\documentclass[12pt,preprint]{aastex}

\def\alwaysmath#1{\ifmmode{#1}\else{$#1$}\fi}

\begin{document}

\title{A 2MASS All-Sky View of the Sagittarius Dwarf Galaxy: 
III. Constraints on the Flattening of the Galactic Halo}

\author{
Kathryn V. Johnston\altaffilmark{1},
David R. Law\altaffilmark{2,3}, and 
Steven R. Majewski\altaffilmark{3}
}

\altaffiltext{1}{Wesleyan University, Department of Astronomy, Middletown, CT
(kvj@astro.wesleyan.edu)}

\altaffiltext{2}{California Institute of Technology, Department of Astronomy, MS 105-24,
Pasadena, CA 91125 (drlaw@astro.caltech.edu)}

\altaffiltext{3}{Dept. of Astronomy, University of Virginia,
Charlottesville, VA 22903-0818 (srm4n@virginia.edu)}

\begin{abstract}
M giants selected from the Two Micron All Sky Survey (2MASS) 
have been used to trace streams of tidal debris apparently associated 
with the Sagittarius dwarf spheroidal galaxy (Sgr) that entirely 
encircle the Galaxy.  While the Sgr M giants are generally aligned with a single 
great circle on the sky, we measure a difference of 10.4 $\pm$ 2.6 degrees 
between the mean orbital poles of the great circles that best fit debris leading 
and trailing Sgr,
which can be attributed to 
the precession of Sgr's orbit over the range of phases explored by 
the data set.
Simulations of the destruction of Sgr in potentials containing bulge, 
disk and halo components best reproduce this level of 
precession along the same range of orbital phases if the potential contours of the
halo are only slightly flattened, with 
the ratio between the axis length perpendicular to and in the disk in the range
$q=0.90-0.95$
(corresponding to isodensity contours with $ q_\rho\sim$ 0.83 - 0.92).
Oblate halos are strongly preferred over prolate ($ q_\rho>1$) halos, and
flattenings in the potential of  $q \le 0.85$ ($q_\rho \le$ 0.75) 
and $q \ge 1.05$ ($q_\rho \ge$ 1.1) are 
ruled out at the 3-$\sigma$ level. 
More extreme values of $q \le 0.80$ ($q_\rho \le$ 0.6) and $q \ge$ 1.25 ($q_\rho \ge$ 1.6) are ruled out
at the 7-$\sigma$ and 5-$\sigma$ levels respectively.
These constraints will improve as debris with larger separation in orbital 
phase can be found.
\end{abstract}


\section{INTRODUCTION}



Simulations of structure formation within the standard Cold Dark Matter
cosmology predict that cluster-scale dark matter 
halos  should typically be far from spherical, with 
shortest to longest axis ratios in density typically around $(c/a)_\rho=0.5$,
rarely greater than $(c/a)_\rho=0.8$, and with no strong preference for oblate ($q_\rho=(c/a)_\rho<1$)
or prolate ($q_\rho=(a/c)_\rho>1$) halos
\citep{dubinski94,jing02}.
Such studies have typically been limited to studying the shapes of halos on cluster-scales
because the computational expense of resolving a large enough
sample of galaxy-scale halos was prohibitive.
Indeed, observations leading to estimates of $(c/a)_\rho$ for a handful of 
external galaxies proved to be roughly consistent with these findings (see Merrifield 2002, Fig. 3 
for a summary).
More recently, preliminary analysis of {\it galaxy}-scale dark matter halos
extracted from cosmological simulations predict that they could be
systematically rounder than their larger counterparts, 
peaking around $(c/a)_\rho=0.65$ and with examples as round as $(c/a)_\rho=0.95$
\citep{bullock02,flores04}, 
and \citet{kazant04} have demonstrated that simulations which
include gas cooling effects also produce rounder halos than similar simulations which
neglect such dissipation.
This result is especially intriguing given at least one 
estimate for the shape of the Milky Way's dark
matter halo that suggests that it could indeed be rather spherical  \citep[$(c/a)_\rho>0.95$,][]{ibata01},
placing it at the extreme of the extragalactic $(c/a)_\rho$ range
\citep[but cf.][who find $(c/a)_\rho \approx$ 0.5]{martinez04}.
However, this estimate was 
made using just a few dozen carbon stars thought to be associated with tidal debris from 
the Sagittarius (Sgr) dwarf galaxy because of their alignment with its orbit on the sky (velocity 
measurements and distance estimates could not conclusively support this
association in most cases), and hence the accuracy of this measurement remains unclear.

Debris from the destruction of a satellite galaxy can provide a sensitive probe of 
deviations of the potential of the Milky Way from spherical symmetry.
Such debris occupies orbits with a range of azimuthal time periods about the 
satellite's own, and this leads to the phase mixing of debris ahead and behind the 
satellite along its orbit  to form tidal tails \citep{johnston98}.
\citet{helmi99} showed that in spherical potentials these tidal tails will 
gradually thicken within the orbital plane, due to the range of precession 
rates of turning points in debris orbits.  In non-spherical potentials, the 
range in precession rates of the orbital poles will lead in addition to 
thickening 
of the debris perpendicular to the (instantaneous) orbital plane 
of the satellite. Hence, if the Milky Way is close to spherical, debris should 
always remain planar and appear close to a single great circle in the sky 
\citep{johnston96}, but if the Milky Way is non-spherical
the debris should, over 
time, spread to cover
a significant fraction of the sky.
Such an alignment of a sample of faint, high latitude carbon stars
\citep{totten98} along a single great circle on the sky previously led \citet{ibata01} to conclude  
$(c/a)_\rho > 0.95$.  


More recently, \citet{helmi04a} has 
cautioned that, since it can take a few orbits for debris to spread 
beyond its intrinsic thickness, stars released only a few orbits ago 
cannot be sensitive probes of potential flattening, and hence
that none of the Sgr data sets thus far detected provide any significant 
constraint on $q$.
In addition to Helmi's (2004a) concern,
the thickness of a debris stream will also increase with 
satellite mass. Hence, if either the current mass or recent mass loss rate is poorly
known these introduce additional uncertainties
in trying to measure $q$ from the thickness of a stream.
As a final concern (and one that is much harder to correct for)  thickening can also result from
interactions with other massive --- and possibly invisible --- lumps in the halo
\citep{moore99,ibata02,johnston02}. 

In this paper we revisit the question of what constraint tidal tail 
stars associated with the Sgr dwarf can place on the 
flattening of the Galactic potential.  Our approach differs 
significantly from previous work in that we use the orbit-alignment rather
than thickening  of the debris to measure the halo shape.
For our analysis, we use M giants 
selected from the Two Micron All Sky Survey (Majewski et al, 2003 --- hereafter Paper I).
The Sgr M giants are thought to have ages of 2-3 Gyr (Paper I) and hence, the {\it thickness} 
of tidal streams constituted of these stars
cannot be expected to provide a strong constraint on $q$ (even ignoring the additional 
concerns about this method we raised above).
However, the approximate distances estimated from the apparent magnitude of the 
M giants permit a clear separation between stars leading or trailing 
Sgr along its orbit (because continuous streams of debris can be traced 
directly back to the Sgr core) and, moreover, yield a detailed view of the 
three-dimensional configuration of these tidal tails (e.g., Paper I).  
As a consequence, rather than 
simply looking at the thickness of debris projected on the sky, we are 
able to measure the poles of the best-fit great circles (in two dimensions) or 
best-fit planes (in three dimensions) of the 
leading and trailing debris separately.  We use the difference in orbital poles between 
the two debris trails to quantify the precession of the orbital plane over this range 
of orbital phase.  The advantage of this powerful approach is that what we are 
measuring is sensitive to $q$ alone --- the thickness of the debris 
due to parent satellite mass, debris age, or scattering off of lumps in the halo
potential do not have systematic effects in this signal and merely contribute 
to the uncertainty in our measurement.\footnote{It is conceivable that 
internal rotation of Sgr perpendicular to the orbital motion could
lead to a systematic offset between the angular momentum distributions of
the leading/trailing debris, mimicing orbital precession, but we do not consider this
since no systematic rotation in Sgr's core has been observed.}  

To determine the expected variation in orbital poles for different values of $q_\rho$, 
we run a series of N-body simulations that mimic the Sgr system as traced by M giants.
We describe our numerical simulations in \S2, compare them to the 
M giant data set in \S 3 and summarize our results and discuss future 
prospects for this approach in \S4. 

\section{METHODS}


Our simulation technique closely follows that outlined in \cite{johnston96}. The 
Milky Way is represented by a smooth, rigid potential, and Sgr by a collection 
of $10^5$ self-gravitating particles whose mutual interactions are calculated 
using a self-consistent field code \citep{hernquist92}.

As a starting point for our study we use the results of Law, Johnston, 
\& Majewski (2004 --- hereafter referred to as Paper IV; preliminary results 
were presented in Law et al. 2004),
 who perform simulations of satellite disruption along orbits
 that are consistent with Sgr's current position, 
line-of-sight velocity and direction of motion tangential to the line-of-sight
(deduced from the orientation of the M giant plane).
The mass and amplitude of tangential motion ($v_{\rm tan}$)
of Sgr and the potential of the Milky Way are systematically 
varied to find the closest possible fit (hereafter referred to as the default model)
to the angular position, distance, and available line-of-sight velocity data for the 
2MASS M giants along the leading/trailing streams. 
The default model is run in a three-component model 
for the Galactic potential, consisting of a Miyamoto-Nagai (1975) disk, Hernquist 
spheroid, and a logarithmic halo:
\begin{equation}
        \Phi_{\rm disk}=-\alpha {GM_{\rm disk} \over
                 \sqrt{R^{2}+(a+\sqrt{z^{2}+b^{2}})^{2}}},
\label{diskeqn}
\end{equation}
\begin{equation}
        \Phi_{\rm sphere}=-{GM_{\rm sphere} \over r+c},
\label{bulgeqn}
\end{equation}
\begin{equation}
        \Phi_{\rm halo}=v_{\rm halo}^2 \ln (R^{2}+(z^2/q^2)+d^{2}),
\label{haloeqn}
\end{equation}
where $\alpha =$ 1, $M_{\rm disk}=1.0 \times 10^{10}$ $M_{\odot}$, 
$M_{\rm sphere}=3.4 \times 10^{10}$ $M_{\odot}$, $v_{\rm halo}= 114$ km s$^{-1}$,
$a=6.5$ kpc, $b=0.26$ kpc, $c=0.7$ kpc, and $d=12.0$ kpc and $q=1$ (i.e. the halo 
component is spherical in the default model).
The distance of the Sun from the Galactic Center is taken to be $R_{\odot} =$ 7 kpc.
Initially, the particles in the Sgr model are distributed 
according to a Plummer (1911) model
\begin{equation}
        \Phi=-{GM_{\rm Sgr,0} \over \sqrt{r^2+r_{\rm 0}^2}},
\label{PlummerEqn}
\end{equation}
where $M_{\rm Sgr,0}=7.5 \times$ $10^8$ $M_{\odot}$ is the initial mass of Sgr and 
$r_{\rm 0}=$ 0.82 kpc is its scale 
length. The satellite is allowed to evolve over five pericentric passages along 
an orbit with a pericenter of 14 kpc, an apocenter of 58 kpc, and a radial orbital 
time period of 0.9 Gyrs. (The exact extent of the orbit depends on the distance 
scale adopted for the M giants, which is currently set by assuming a 24 kpc 
heliocentric distance for the Sgr core.)   At the point matching Sgr's current 
position and line-of-sight velocity, the satellite in the default model has a 
bound mass of $3\times 10^8M_\odot$. In Paper IV we find that the data still allow 
some freedom in the exact form for the Galactic potential, but that within a 
given potential the final bound mass of Sgr
is constrained to within a factor of two, and, once this
is known, $v_{\rm tan}$ can be found to within a few km s$^{-1}$. 


Paper IV, in effect, finds the best-fitting set of parameters that describe the configuration
of Sgr debris {\it projected onto} the instantaneous orbital plane, whereas the current discussion
focuses on the variations in the position of the orbital plane itself.
In this paper, we rerun Paper IV's default model with a variety of halo flattenings
ranging from $q =$ 0.8 to 1.45 (i.e., we consider both oblate and prolate potentials).
Since we are interested in constraining the contours of the potential, we flatten these directly
via parameter $q$ in Equation (\ref{haloeqn}) rather than generating them from a flattened
density distribution.  These oblate/prolate potentials correspond to even greater
oblateness/prolateness (e.g. Fig. 1 in Helmi 2004a) of the isodensity contours in the effective
distance range that Sgr's orbit explores.  Note that even our models with spherical halos have
modestly flattened potential contours due to the contribution of the Galactic disk.


We recalculate the values of the Galactic halo scalelength $d$ and
$v_{\rm tan}$ for each value of $q$ considered to obtain the best possible fit to the M giant
distance and line-of-sight velocity data within each model of the Galactic potential (see Paper IV
for a description of this fitting process).  The calculated values for each choice of $q$ are given
in Table 1.  Note that the orbital
characteristics of the model dwarf will vary slightly for each of these models, 
and therefore the mass-loss history of the dwarf will depend on $q$.
However, we do not revisit the question of the best-fit 
mass for Sgr in each of these potentials because we
are only concerned with the differential precession of the poles of the tidal debris, which is
independent of satellite mass.

Throughout the paper we deliberately compare the M giant data with our
simulations as viewed from the Sun, despite the fact that orbital precession is most
naturally discussed in reference to the Galactic Center. This is because
transforming the data to a Galactocentric viewpoint would introduce
distortions due to uncertainties in the distances to both the M giants 
and to the Galactic Center that might be confused with orbital precession.
In contrast, our heliocentric viewpoint of the simulations is fixed by only requiring that
our simulated Sgr has the correct current position and velocity relative to the Sun. 
Moreover, we prefer a heliocentric system because is the natural system for the
Great Circle Cell Count analysis we perform in \S 3.2.

\section{RESULTS}


%

In \S3.1 we present a qualitative assessment of Aitoff projections of Sgr debris, and demonstrate
similarly to previous discussions (e.g. Helmi 2004a) that such qualitative analyses provide little
information on halo flattening.  In contrast,
we adopt two quantitative approaches to assess the degree of orbital plane precession of the Sgr
system.  The first method (\S3.2) relies on careful measurements of the mean great circle
described by the $(l,b)$ distributions of M giants in the leading and trailing tidal tails.  As applied here,
this method is virtually free of observational errors or errors in derived quantities, particularly
the derived distances of the M giants.  In the second method (\S3.3) we analyze the 
orbital plane variation using the derived three dimensional distribution of the Sgr M giants.
Both techniques advance previous discussions of the qualitative appearance of Sgr debris in Aitoff 
projections to show that a slightly oblate Milky Way potential is preferred.


\subsection{Aitoff Projections}

Figure \ref{xyplot} shows the projection of the final positions of 
particles in the default simulation onto Sgr's orbital
plane, and demonstrates the $\Lambda_{\odot}$ coordinate system of Paper I; $\Lambda_{\odot}$ is
the angular distance from 
Sgr along the plane of its orbit, defined to be 
zero at the core and increase in the trailing direction.
The colors 
represent different debris ``eras'', i.e. orbits (denoted as one 
apoGalacticon to the next apoGalacticon) on which the debris was 
stripped from the satellite --- yellow for particles lost since 
apoGalacticon about 0.4 Gyr ago, magenta, cyan, and green for particles 
stripped from the dwarf two, three, and four orbits ago, respectively (the orbital
period of Sgr in all of our model Galactic potentials is about 0.75-1.0 Gyr).  
In Paper IV we find that the positions and available velocities for Sgr M giants
presented in Paper I are consistent with debris younger than
about 1.9 Gyr (i.e. yellow, magenta or cyan) in these simulations.
This age estimate is consistent with the expected 
2-3 Gyr lifetimes of these stars --- for the M giants to be a significant contributor to debris older
than about 1.9 Gyr
the stars would have to be stripped from the satellite less than one orbital period following their birth 
(see Paper I for a fuller discussion of this problem). 

Figure \ref{aitoffplot} plots Aitoff projections of the final positions of particles 
in both spherical and moderately flattened halos ($q=1$ and $q=0.85$).
Visual inspection of this figure confirms the \citet{helmi04a} conjecture that 
{\it qualitative} differences in the apparent thickening or precession 
of the most recent debris  (i.e. yellow, magenta, and cyan points) 
are fairly small over the range of 
potential flattenings shown, especially for younger debris.  
However, it is premature to conclude that these
effects do not lead to measurable variations in the disposition 
of the Sgr debris.  We now demonstrate that {\it orbital precession} of
even the most recent Sgr debris can be measured and used to discriminate
between halo flattening models.

\subsection{Great Circle Cell Orbital Poles} 

In their current configuration about the Milky Way, Sgr leading arm M giants 
predominate in the Northern Galactic Hemisphere while trailing debris M giants
dominate the South.  Paper I 
demonstrated (see their Fig. 6) 
clear differences in the location of the Sgr M giant Great Circle Cell Count
\citep{johnston96}[``GC3"] peaks when the M giant sample is limited to $b>+30^{\circ}$
versus $b<-30^{\circ}$ subsamples, and suggested that the differences
could relate to a combination of precessional and parallax differences between
leading and trailing arm debris.  We repeat the Paper I GC3
analysis on our simulations (using the yellow, magenta, and cyan debris that best represent
the Paper I M giant data) 
to determine the degree of precessional shift
as a function of $q$.  Figure \ref{gc3plot} compares the shape and positions of the 
GC3 peaks of the models to those of the M giant data, divided into
$b>+30^{\circ}$ (North/mostly leading arm) and $b<-30^{\circ}$ (South/mostly
trailing arm) subsamples.  Other aspects of sample selection, including limiting
to stars with distances between 13 and 65 kpc and ignoring the region around the
Magellanic Clouds ($260^{\circ}<l<320^{\circ}$ and $-53^{\circ}<b<-25^{\circ}$) were
utilized to exactly match the selection critria used to generate the GC3 plots in
Paper I (their Fig.\ 6).  The 13-65 kpc distance range limit serves
to highlight the primary distance range for the
debris corresponding to the yellow, magenta and cyan parts of the leading and trailing
arms.  For the observations, this limit also removes nearby disk M giants and 
more distant ``M giants" at magnitudes where 2MASS
photometry becomes less reliable, whereas for both the observations and the models it
limits the effects of overlapping contaminating debris 
on each Sgr arm from extensions of the opposite Sgr arms.
The shapes of GC3
peaks reflect departures from great circle symmetry {\it as viewed from the Sun}, 
while the size (e.g.,
FWHM) of the peak is a function of the debris width on the sky, convolved with 
the cell size (here adopted as a 5$^{\circ}$ wide cell\footnote{Runs with narrower cell 
widths yield virtually identical peak locations and shapes to the results for 5$^{\circ}$ wide
cells presented here, albeit with poorer signal-to-noise because of the fewer numbers 
of stars that fall within the narrower cells.}). The first thing
to notice is the overall consistency of the simulated and actual M giant data in both 
the shape and size of the GC3 peaks for {\it both} hemispheres.  The detailed 
matches are encouraging support for the conjecture that the 
simulations provide rather accurate representations of the actual Sgr 
M giant distribution. 

The second thing to note is that while there is virtually no difference in 
the location of the GC3 peaks for the southern (trailing arm) data as a function of
$q$, the position of the GC3 peak for northern (leading arm) data is, in contrast,
highly and systematically sensitive to $q$. 
The effect is more pronounced in the North
because (a) the leading arm extends very close to (virtually on top of) the Sun and a 
given precessional shift relative to the Galactic center
will be foreshortened to a larger angular shift on the sky for closer debris, and (b) the northern
sample utilized includes a larger fraction of {\it older} debris, torn off from 
Sgr one orbit earlier (i.e., the cyan debris in Fig.\ 1), and these very stars, which
will have experienced more precession, constitute
much of the Sgr debris closest to the Sun in the North.
Note that the simulations
were not adjusted to create a best match to the GC3 poles in Figure \ref{gc3plot}, but rather
the orbit of Sgr in the simulations was set by the mean pole 
([$l,b$] = [273.75,-13.46]$^{\circ}$) found in the Paper I single plane 
fit to the Cartesian positions of both leading and trailing M giant debris; 
the close match to the 
general positions of the separated North/South GC3 poles arises naturally 
from the evolution of the debris.\footnote{Whereas a more 
precise match may be possible through trial and error adjustment
of the instantaneous orbital pole of the Sgr core, this exact match is not essential
to our goal of measuring {\it differences} between the leading and trailing debris.} 
The poles in Figure \ref{gc3plot} show the simulations with $q=0.90$ and $q=0.95$ 
to be the closest match to both the absolute positions of the northern
hemisphere GC3 peak as well as the relative
difference in North/South GC3 peak positions in the M giant data.
More oblate and all prolate models yield northern GC3 peaks that
are tens of degrees off from the observed positions.  The GC3 analysis
shows that subtle variation in the $(l,b)$ distribution of debris (Fig.\ 2)
{\it can} be measured and, in a way virtually free of observational bias, be
used to constrain $q_\rho$.


\subsection{Plane Fitting Poles}


The GC3 analysis gives a phase-averaged view of the expected differential precession of Sgr
leading and trailing debris.
In contrast, the lines in Figure \ref{poleplot}
show the pattern that the {\it instantaneous} pole of Sgr's orbit traces on the 
sky within $\pm$1.5 Gyrs of Sgr's current position, as viewed from the Sun.
The upper/lower panels show the pole evolution in potentials with oblate/prolate
halo components respectively, with the solid/dotted portions of each line corresponding to
portions of the orbit trailing/leading Sgr. The black 
triangle shows the best fit pole of the 2MASS data set (Paper I), which the orbits 
were constrained to go through at Sgr's current position.

The dots along each curve in Figure  \ref{poleplot}
indicate the pole position at 0.2 Gyr intervals both leading 
and trailing Sgr's current location.
The M giants discussed in Paper I explore orbital phases corresponding to up to (roughly) +0.6 Gyrs
and -0.4 Gyrs from Sgr's current position, ahead and behind 
it in time along the orbit.
(Note that the debris age --- i.e. the time since stars were lost from Sgr --- is much greater than
this.)
Hence, the poles actually measured from the M giant 
and simulation data can be thought of as averages of 3 points along the dotted curves
and 2 points along the solid curves, 
weighted by the density of stars or particles along the streamers.
(Note that, for fixed flattening, the pole paths of orbits with $v_{\rm tan}$ varying from 
the adopted value by as much as 
$\pm$10 km s$^{-1}$  --- more than the maximum range permitted given
constraints from the M-giant distances and velocities, as shown in Paper IV ---
were found to be virtually  indistinguishable from the orbits illustrated.).
This figure demonstrates  (as already seen in Fig. \ref{gc3plot}) that:
(i) In general, the separation of poles derived for the leading and trailing simulated data
should increase with increasing deviations from $q=1$;
(ii) pole separation should be more dramatic for the oblate cases than the prolate cases;
(iii) the sense of precession in prolate potentials is opposite to that in oblate potentials;
and (iv) some (oblate-like) precession should be present even in the $q=1$ case because 
of the presence of a disk component in our Galactic potential.

We might expect the lines in Figure \ref{poleplot} to move monotonically in Galactic
longitude.
The retrograde-looping
evolution of the leading (dotted) portion in the oblate cases is due to our 
perspective of these orbits, which have pericenters that lie only just 
outside the Solar Circle. Indeed, some
asymmetry between the evolution of the poles of leading and trailing debris 
is already apparent in the more dramatic shift 
of the GC3 peaks of the Northern (leading) debris 
when compared to the Southern (trailing), as noted in
the previous section (Fig. \ref{gc3plot}).
However, the Figure \ref{gc3plot} GC3 analysis is a somewhat blunt tool for assessing
the precessional shifts because it does not take advantage of all 
available information --- namely, the distances to the stars/particles.

To quantify the degree of precession exhibited by Sgr debris better, 
the poles of best-fit orbital planes can be determined for both the M giant and the
simulated data by plane-fitting
to the leading and trailing arm debris in three-dimensional space.
We constrain these planes 
to pass through the solar position to allow us to quantify the effect
precession has on Sgr tidal debris as projected on the sky (the natural
observational regime, e.g., as used in a GC3 analysis).
This constraint means that the poles derived by this method should be interpreted
only as tools for tracing the amount of
precession in a given model of the Galactic potential; they are not the true
Galactocentric orbital poles and do not precisely represent 
the angular momentum vector of the debris.  The best-fit planes to the M giant and 
simulation distributions were found by minimizing the $\chi^2$ distribution of 
point distances from the plane, applying an iterative 2.5-$\sigma$ rejection
algorithm.
Confidence limits on the corresponding poles were determined by
statistical analysis of synthetic data sets generated using a bootstrapped
Monte Carlo technique, with the confidence ellipse resulting from this analysis
projected into error bars at the 68\% confidence level in $l$ and $b$.

\begin{figure}
\epsscale{0.9}
\plotone{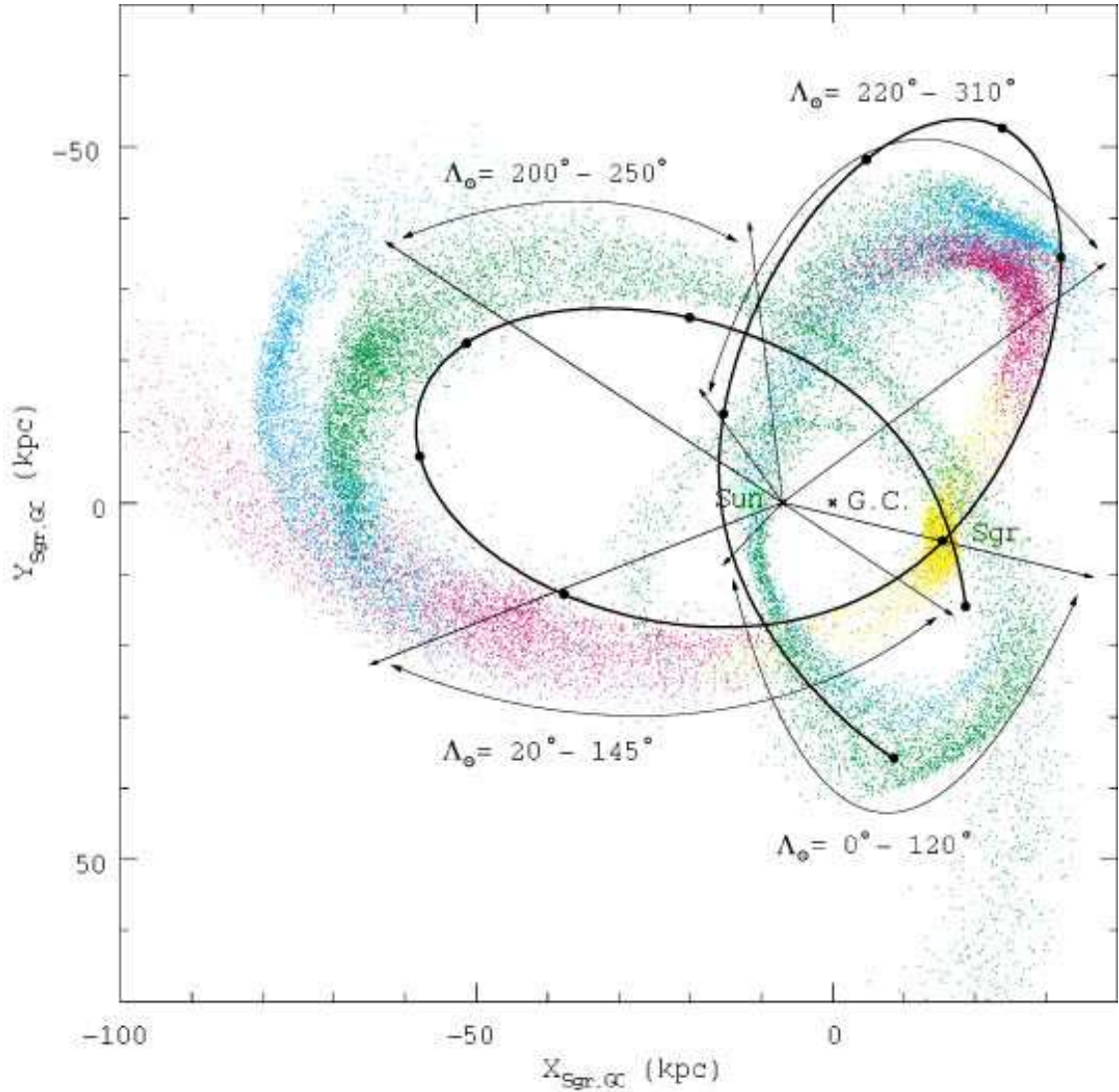}
\epsscale{1.0}
\caption{
Appearance of a typical model of Sgr tidal destruction (from Paper IV) in the
Sgr plane coordinate system (see Paper I).   The location of the
Sun, Galactic Center  and Sgr dwarf are marked.
Yellow particles became unbound from the dwarf
since the last Sgr apoGalacticon (about 0.4 Gyr ago), and particles
lost in successive preceding orbits are coded as magenta, cyan
and green points.  The orbit of Sgr over the last $\pm$1 Gyr is indicated by the bold line, with tickmarks
representing time invervals of 0.2 Gyrs.  Labels indicate regions of
debris used for precession analysis in Section 3.3.}
\label{xyplot}
\end{figure}

\begin{figure}
\epsscale{0.9}
\plotone{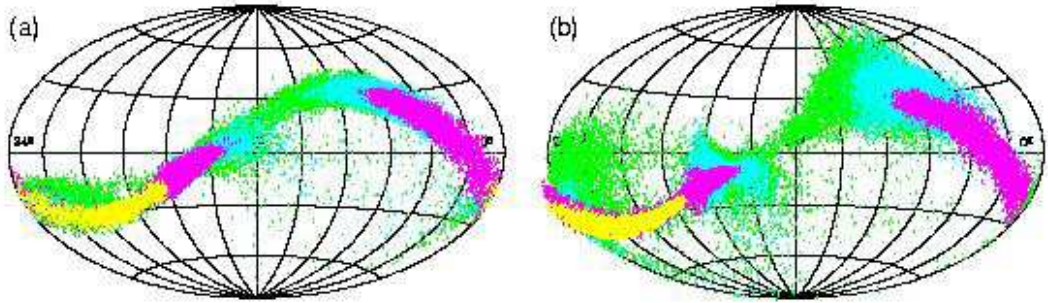}
\epsscale{1.0}
\caption{Aitoff projections of debris particles in our simulations
with (a) $q=1$ and (b) $q=0.85$, with the same color coding as Figure \ref{xyplot}. }
\label{aitoffplot}
\end{figure}

\begin{figure}
\epsscale{0.9}
\plotone{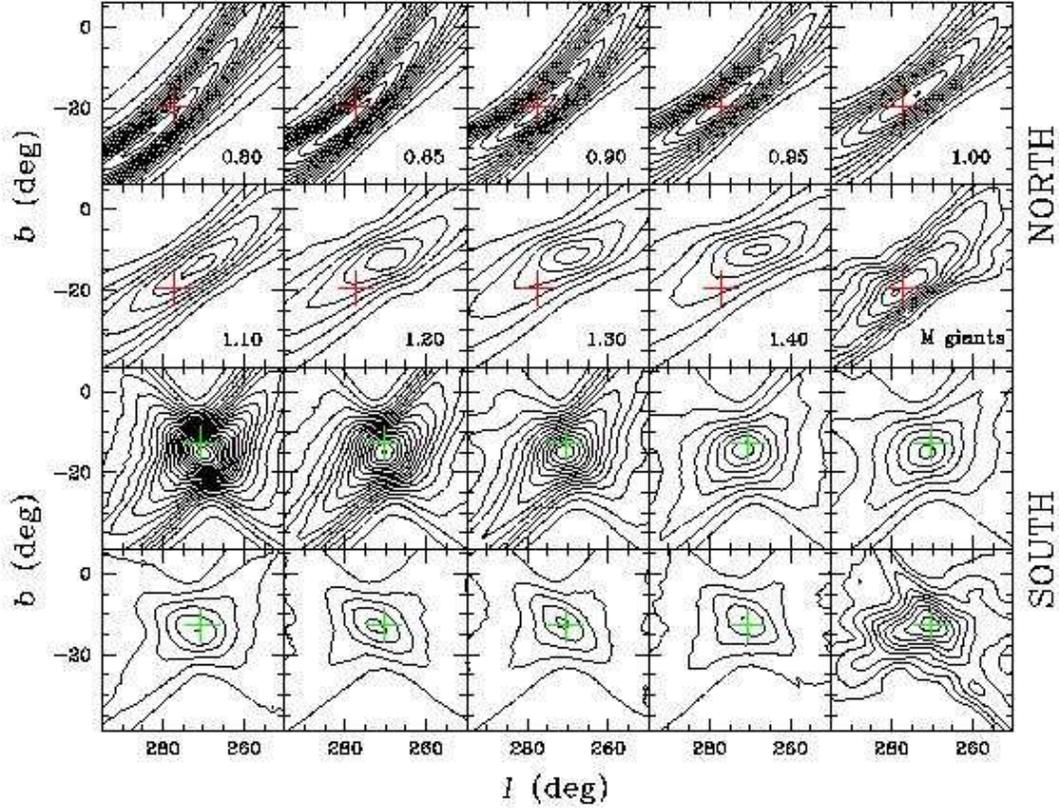}
\epsscale{1.0}
\caption{Close-up views of the 
GC3 peaks corresponding to Sgr debris in simulations with varying
$q$ parameters as well as to the actual M giant data.
The M giant data shown are exactly those in Figure 6 of Paper I.
The upper two rows show the peaks derived from GC3 analysis of
all simulation particles or M giants with $b>+30^{\circ}$ while
the bottom two rows correspond to simulations/data with $b<-30^{\circ}$.
The $q$ values shown in the panels in the upper rows are not repeated in
the bottom panels (for clarity), but map identically from the upper ten to
the lower ten panels.
The contour levels have linear spacing, separated in steps of
400 for the North simulations, 150 for the South simulations, 
and 20 for the actual M giant data.  The resolution of the 
GC3 maps is 1 degree in both $l$ and $b$, while the cell size for the maps shown
are 5$^{\circ}$.
The cross symbols are North (red) and South (green) reference guides, 
set to the approximate centers of the peaks in the North 
and South M giant data respectively, and are repeated in 
the same locations across the maps for the corresponding simulations.  A clear shifting of 
the orbital pole positions with $q$ is evident in the Northern
Hemisphere (leading arm), but to a much less extent in the Southern Hemisphere (trailing
arm).  The simulations that 
match the M giant data closest are those with $q=0.90$ and $q=0.95$.
  }
\label{gc3plot}
\end{figure}

\begin{figure}
\plotone{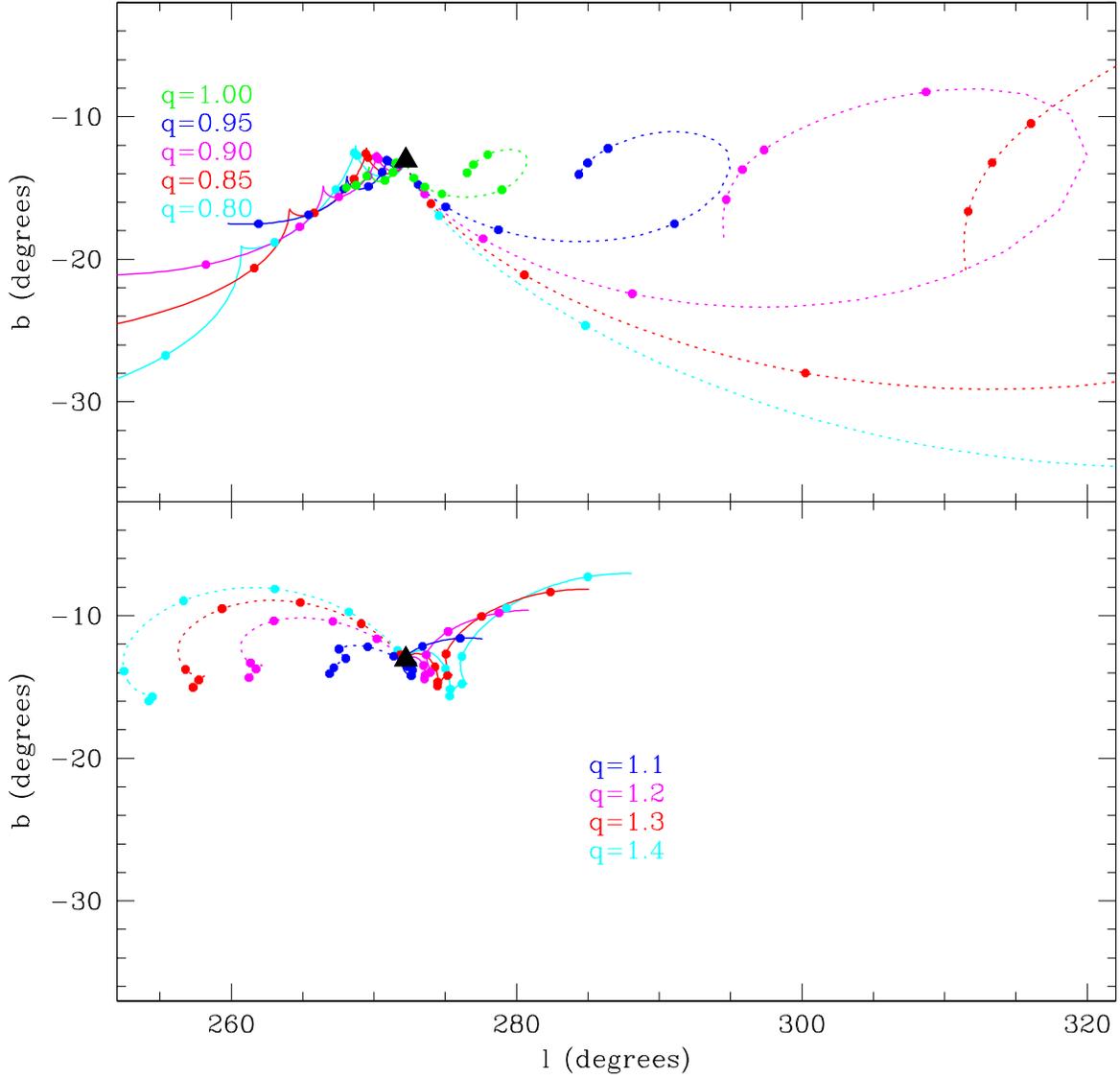}
\caption{The evolution of Sgr's instantaneous orbital pole 
$\pm$1.5 Gyrs along its orbit in potentials with $q\le 1$ (top panel)
and $q > 1$ (bottom panel). The {\it black triangle} shows
the pole calculated from the combined leading and trailing M giant debris. 
Tickmarks on each evolutionary path 
represent time intervals of 0.2 Gyrs and the solid/dotted lines indicate
trailing/leading portions (see text). 
}
\label{poleplot}
\end{figure}

\begin{figure}
\plotone{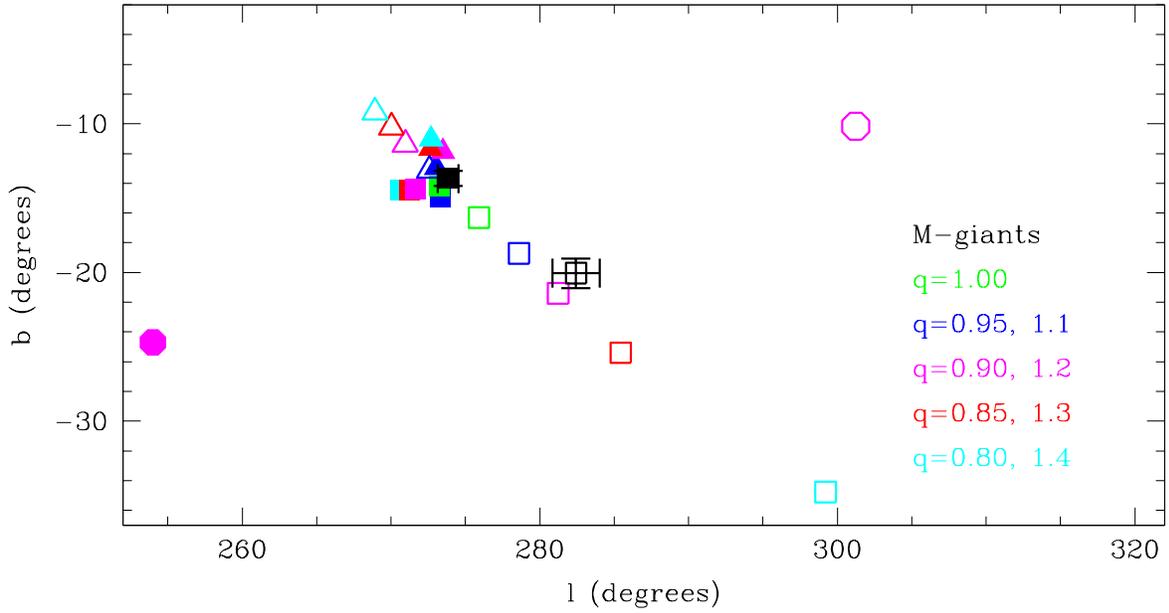}
\caption{The positions of 
apparent orbital poles for the M giants and
simulated debris calculated using the 
plane-fitting technique. 
Open/filled symbols represent the poles of leading/trailing debris,
squares/triangles are for oblate/prolate potentials.
Circles represent orbital poles for the $q =$ 0.90 simulation calculated
from older, more greatly precessed tidal debris.
}
\label{planeplot}
\end{figure}

The open symbols in Figure \ref{planeplot} show the results of this 
precession analysis applied to data points in the leading tail in the 
range $\Lambda_{\odot}=220^{\circ}$ - $310^{\circ}$ , while the filled symbols
show the results for  the trailing 
tail in the range $\Lambda_{\odot}=20^{\circ}$ -$145^{\circ}$ 
\footnote{These
were selected as ranges in which Sgr tail M giants may readily be 
identified in the 2MASS database using the selection criteria $E(B-V) < 0.555$, 
$1.0 < J-K_s < 1.1$, $|Z_{\rm Sgr,\odot}| < 25$ kpc, $|b| > 30^{\circ}$, and 
distances 13 kpc $< d_{\ast} < 60$ kpc.  See Paper I for definitions of these
criteria.
The Magellanic Clouds were removed from 
this data set using the $(l,b)$ cuts given in Paper I. 
For the simulations, yellow, magenta, and cyan debris were considered and
subjected to the same cuts in $\Lambda_{\odot}$, $d_{\ast}$, and $b$
as the M giants.} --- see Figure \ref{xyplot}.
The colored triangles/squares correspond to simulated data run in prolate/oblate potentials.
(The error bars on the simulated points are
comparable to the symbol sizes and are omitted for clarity.)
Note there is a systematic offset in $(l,b)$ of order a few 
degrees between the results from the observed data and the trend of the simulated data. This can 
be attributed to our assumption that the pole derived from the full 2MASS data 
corresponds to Sgr's present orbital pole when, in fact, it merely represents 
an average along tidal debris with a small range of orbital phases.
Nevertheless it is clear that: (i) the {\it sense} of precession in the data strongly favors
oblate potentials; and (ii)
the evolution of pole {\it differences} in the plot 
implies that the precession rate is most consistent with our slightly flattened halo 
models in which $q \approx 0.90$. 
Table 1 emphasizes this result by quoting the separation of the 
leading/trailing poles for the simulations and the M giants. 
Flattenings within the range $0.90<q<0.95$ ($0.83<q_\rho<0.92$)
lie within the 1.5-$\sigma$ error bars on the data, 
while those with $q<0.85$  and $q>1.05$ are ruled out at the
3-$\sigma$ level.
More extreme values of $q \le 0.80$ ($q_\rho \le$ 0.6) and $q \ge$ 1.25 ($q_\rho \ge$ 1.6) are ruled out
at the 7-$\sigma$ and 5-$\sigma$ levels respectively.

The open/filled circles in  Figure \ref{planeplot} show the results of the same analysis 
performed on the older (color coded green in Figs. \ref{xyplot} and \ref{aitoffplot}) simulated data
at larger separations along the orbit 
($\Lambda_{\odot}=0^{\circ}-120^{\circ}$ for the leading debris and 
$200^{\circ}-250^{\circ}$ for the trailing --- see Fig. \ref{xyplot}) for the $q =$ 0.90 simulation, 
which best reproduces the poles of the younger debris. 
The positions of the circles are suggestive of the evolution traced by the orbital 
poles in Figure \ref{poleplot}.
Note that the M giants discussed in Paper I are primarily in the yellow, magenta and cyan portions of 
the tidal debris (see Paper IV) and there is no clear evidence yet for
M giants in the portions of the tails corresponding to the green debris; 
the phase-mixing time for debris to reach these points is comparable to the 
stellar evolution ages of these stars (see \S3.1). 
The circles in Figure \ref{planeplot} demonstrate that an analysis such as that conducted here 
will provide a more powerful constraint on $q$ if tracers of this older 
debris can be found, so long as leading debris, 
which will be mainly in the {\it Southern} Hemisphere
for ``green debris", can clearly be separated 
from the {\it trailing} debris there 
 using distance and/or velocity information. 
Indeed, at these larger phase-differences from Sgr, the 
leading/trailing poles differ by more than 45 degrees even for flattenings
of $q= 0.90$. This suggests that the accurate determination of the 
centroid of an older piece of the Sgr tidal stream, even at a single longitudinal point,
can provide strong leverage on $q$ via comparison with test-particle integrations in 
different potentials.

\section{CONCLUSIONS AND FUTURE PROSPECTS}


We have shown that the precession of Sgr's orbit can be accurately 
traced by stars in its debris streams --- even the relatively 
recently released M giant stars --- and hence be used to constrain 
the flattening of the Galactic potential.  A difference 
of $10.4\pm2.6$ degrees is found between the best-fit orbital poles for stars in Sgr's 
leading and trailing streams (in the sense that the angular momentum vector
{\it increases} in $l$ and {\it decreases} in $b$ from trailing to leading debris) when those
poles are measured from debris
at azimuthal orbital phases within +140/-145 degrees 
along its orbit as viewed from the Sun.
Such a low amount of precession is 
most consistent (within 1.5-$\sigma$) with simulations of the destruction of Sgr run in 
models of the Galaxy with a slightly flattened halo where $q$ is in the range 0.90 - 0.95
($q_\rho =$ 0.83 - 0.92).
Flattenings for the halo potential of $q=0.85$ ($q_\rho=0.75$) or less 
and $q=1.05$ ($q_\rho=1.1$) or more
are ruled out at the 3-$\sigma$ level, and oblate models are strongly preferred over prolate models. 
Note that these results depend on the
assumed form of the potential of the Galactic disk.

Recently, Helmi (2004b) has suggested that prolate Milky Way potentials offer
a means by which to solve a dilemma discussed in Paper IV --- namely that the Sgr leading arm
radial velocities are one discrepant observable not well fit by oblate Milky Way + Sgr
models that provide compelling overall fits to all other available observables (see Paper IV).  While a 
prolate model solves this one problem, as shown here prolate models also introduce a serious
discordance with the observed M giant precession; indeed, prolate models are found to induce
precession in the {\it opposite} direction to that observed.
Since orbital pole precession is almost solely 
sensitive to the shape of the potential, whereas a variety of other effects in addition to the halo shape --- 
e.g., evolution in the Sgr orbit and/or the strength of the Galactic potential --- can conceiveably alter 
the dynamics of debris {\it within} the orbital plane, we are inclined to the simpler explanation
of an oblate potential to match the precessional data while admitting the need for yet more 
sophisticated models to resolve the problem with the leading arm velocities using these other
free parameters.

Future detections of older 
Sgr debris at larger phase-differences along its orbit will provide 
even stronger orbital precession constraints than obtained here.
Recently, \citet{newberg03} announced a new detection of Sgr 
debris as an over-density of A-colored stars in the Sloan Digital 
Sky Survey at sufficient angular separation from Sgr 
($\Lambda_\odot \sim 196^{\circ}$) and distance from the Sun ($>80$) kpc to 
possibly associate it with the oldest (green) debris in our simulations. 
Unfortunately, the debris has not yet been mapped fully enough for   
a determination of its mean position; when done, however,
a more accurate determination of $q$ may be possible. 
Thus, Sgr has still more to contribute to our understanding 
of the Galactic potential.



\acknowledgments
SRM acknowledges support from a Space Interferometry Mission Key 
Project grant, NASA/JPL contract 1228235, NSF grant AST-0307851, 
a David and Lucile Packard Foundation Fellowship, and 
the F.H. Levinson Fund of the Peninsula Community Foundation.
KVJ's contribution was supported 
through NASA grant NAG5-9064.and NSF CAREER award AST-0133617.

\clearpage

\begin{table}[h]
\begin{tabular}{c|c|c|c}
\hline
$q$	& $d$ & $v_{\rm tan}$ &	Pole Separation	\\
	& (kpc) & (km s$^{-1}$)	&	(degrees)			\\
\hline
\hline
\multicolumn{3}{c|}{Sgr M giants} &10.4 $\pm$ 2.6 	  \\
\hline
\hline
0.80 & 13 & 286 & 32.7 $\pm$ 0.5\\
\hline
0.85 & 13 & 284 & 17.2 $\pm$ 0.5\\
\hline
0.90 & 13 & 280 & 11.5 $\pm$ 0.5\\
\hline
0.95 & 12 & 272 & 6.3 $\pm$ 0.5\\
\hline
1.00 & 12 & 270 & 3.3 $\pm$ 0.5\\
\hline
1.05 & 12 & 268 & 1.2 $\pm$ 0.4\\
\hline
1.10 & 12 & 264 & 0.5 $\pm$ 0.5\\
\hline
1.15 & 12 & 262 & -2.2 $\pm$ 0.4\\
\hline
1.20 & 11 & 256 & -2.5 $\pm$ 0.4\\
\hline
1.25 & 11 & 254 & -2.9 $\pm$ 0.4\\
\hline
1.30 & 11 & 252 & -2.9 $\pm$ 0.4\\
\hline
1.35 & 11 & 250 & -3.9 $\pm$ 0.4\\
\hline
1.40 & 11 & 250 & -4.1 $\pm$ 0.5\\
\hline
1.45 & 11 & 250 & -5.0 $\pm$ 0.4\\
\hline
\end{tabular}
\caption{Orbital pole precession of tidal debris between
$20^{\circ} < \Lambda_{\odot} < 145^{\circ}$ (trailing debris) and $200^{\circ} < \Lambda_{\odot} < 320^{\circ}$ 
(leading debris) for Sgr M giants
and simulations with indicated values of the Galactic halo flattening $q$.  
Positive separation values indicate simulations whose leading debris orbital pole is at higher/lower l/b than the trailing debris pole,
and negative values indicate simulations whose leading debris pole is at lower/higher l/b than the trailing debris pole
(i.e. in the direction opposite to that observed for Sgr M giants, see Fig. 5).
The values assumed for the halo
scale length ($d$) and the velocity of the model dwarf tangential to the line of sight ($v_{\rm tan}$) 
in each model are also given.}
\end{table}

\clearpage

\end{document}